\documentclass{kybernetika}

\usepackage{amsfonts}
\usepackage{latexsym}
\usepackage{mathrsfs}
\usepackage{verbatim}
\newtheorem{theorem}{Theorem}[section]
\newtheorem{lemma}[theorem]{Lemma}

\newtheorem{definition}[theorem]{Definition}

\newenvironment{proofi}[1][Proof]{\textbf{#1.} }{ \hfill\rule{0.5em}{0.5em}}

\newcommand{\vp}{\mathbf{p}}

\begin{document}
\pagestyle{myheadings}

\title{Universally Typical Sets \\
for Ergodic Sources of Multidimensional Data}

\author{Tyll Kr\"uger, Guido Mont\'ufar, Ruedi Seiler, and Rainer Siegmund-Schultze}

\contact{Tyll}{Kr\"uger}{Universit\"{a}t Bielefeld, Fakult\"{a}t f\"{u}r Physik, Universit\"{a}tsstra\ss e 25, 33501 Bielefeld, Germany}{tkrueger@physik.uni-bielefeld.de}
\contact{Guido}{Mont\'ufar}{Pennsylvania State University, Department of Mathematics, 218 McAllister Building, University Park, PA 16802, USA \\
Current address: Max Planck Institute for Mathematics in the Sciences,  Inselstra\ss e 22,  04103 Leipzig, Germany 
}{gfm10@psu.edu}
\contact{Ruedi}{Seiler}{Technische Universit\"at Berlin, Institut f\"ur Mathematik MA 7-2, Stra\ss e des 17.~Juni 136, 10623 Berlin,
Germany}{seiler@math.tu-berlin.de}
\contact{Rainer}{Siegmund-Schultze}{Universit\"{a}t Bielefeld, Fakult\"{a}t f\"{u}r Physik, Universit\"{a}tsstra\ss e 25, 33501 Bielefeld, Germany}{siegmund@math.tu-berlin.de}

\markboth{T. Kr\"uger, G. Mont\'ufar, R. Seiler, and R. Siegmund-Schultze} {Universally Typical Sets of Multidimensional Data}

\maketitle

\begin{abstract}
We lift important results about universally typical sets, typically sampled sets, and empirical entropy estimation in the theory of samplings of discrete ergodic information sources from the usual one-dimensional discrete-time setting to a multidimensional lattice setting. 
We use techniques of packings and coverings with multidimensional windows to construct sequences of multidimensional array sets which in the limit build the generated samples of any ergodic source of entropy rate below an $h_{0}$ with probability one and whose cardinality grows at most at exponential rate $h_{0}$. 
\end{abstract}



\section{INTRODUCTION}

An entropy-typical set is defined as a set of nearly full measure consisting of output sequences the negative log-probability of which is close to the entropy of the source distribution. The scope of this definition is revealed by the asymptotic equipartition property (AEP), which was introduced by McMillan~\cite{McMillan:1953} as the convergence in
probability of the sequence $-\tfrac{1}{n}\log \mu (x_{1}^{n})$ to a constant $h$, namely, the Shannon entropy rate of the process $\mu$~\cite{ShannonOriginal}. 
Many processes have the AEP, as has been shown, e.g., in~\cite{Bj:04,Breiman:1957,McMillan:1953,OrWe83}. 
In particular, for stationary discrete-time ergodic processes, this property is guaranteed by the Shannon-McMillan (SM) theorem~\cite{McMillan:1953} and in the stronger form of almost-sure convergence by the Shannon-McMillan-Breiman (SMB) theorem~\cite{Breiman:1957}. 
These two theorems have been extended from discrete-time to amenable groups, including $\mathbb{Z}^{d}$ as a special case, by Kieffer~\cite{Kieffer75} and Ornstein-Weiss~\cite{OrWe83}, respectively. 
\newline

Roughly speaking, the AEP implies that the output sequences of a random process are typically confined to a `small' set of events which have all approximately the same probability of being realized, in contrast to the much larger set of all possible output sequences. This means that individual outcomes with much higher or smaller
probability than $e^{-n h}$ will rarely be observed. 
By the AEP, the entropy-typical sets have total probability close to one and their cardinality is fairly minimal among all sets with this property. 
This way, entropy-typical sets provide an important theoretical framework for communication theory. 
Lossless source coding is a type of algorithm which performs data compression while ensuring that the exact reconstruction of the original data is possible from the compressed data. 
Lossless data compression can be achieved by encoding the typical set of a stochastic source with fixed length block codes of length $n h$. 
By the AEP, this length $n h$ is also the average length needed. Hence compression at an asymptotic rate equal to the entropy rate is possible. 
This rate is optimal, in view of Shannon's source coding theorem~\cite{ShannonOriginal}. 
\newline

In universal source coding, the aim is to find codes which efficiently compress down to the theoretical limit, i.e., the entropy rate, for any ergodic source without a need to be adapted to the specific source. 
We emphasize here that codes of that type are optimal data compressors for any \emph{stationary }source, since by the ergodic decomposition theorem (see, e.g., \cite{KSchm}) any stationary source is a convex mixture of ergodic sources. 
Moreover, any asymptotically optimal universal compression scheme defines sequences of universally typical sets: for given $\varepsilon $, the set of all $n$-blocks such that their  compression needs at most $(h+\varepsilon)n$ bits, is universally typical for all sources with entropy rate $h$ or less. 
Vice versa, any \emph{constructive }solution to the problem of finding universally typical sets yields an universal compression scheme, since the index in the universally typical set is an optimal code for the block. 
As will turn out, our approach for multidimensional sources is constructive. But one has to admit that such an \emph{ad hoc }algorithm is, generally speaking, not very useful in practice, because determining the index should be very time consuming.
\newline

Many formats for lossless data compression, like ZIP, are based on the implementation of the algorithms proposed by Lempel and Ziv (LZ) LZ77~\cite{LZ77} and LZ78~\cite{LZ78}, or variants of them, like the Welch modification~\cite{LZW}. 
The LZ algorithms allow to construct universally typical libraries. 
However, they are designed as text compression schemes, i.e., for $1$-dimensional data sources. 
Lempel and Ziv~\cite{2DLZ} showed that universal coding of images is possible by first transforming the image to a $1$-dimensional stream (scanning the image with a Peano-Hilbert curve, a special type of Hamilton path) and then applying the $1$-dimensional algorithm LZ78. 
The idea behind that approach is that the Peano-Hilbert curve scans hierarchically complete blocks before leaving them, maintaining most local correlations that way. 
In contrast, a simple row-by-row scan
only preserves horizontal correlations. 
But with the Peano curve approach, while preserving local correlations in any non-horizontal direction, these correlations are much encrypted due to the inevitably fractal nature of that space-filling curve.
\newline

We take the point of view that the techniques of packing and counting can be better exploited in data compression with unknown distributions if, instead of transforming the `image' into a $1$-dimensional stream by scanning it with a curve, the multidimensional block structure is left untouched. 
This will allow to take more advantage of multidimensional correlations between neighbouring parts of the data, speed up the convergence of the counting statistics, and in turn fasten estimation and compression. 
This approach will be carried out in a forthcoming paper. 
The idea of the present paper is to extend theoretical results about typical sampling sets and universally typical sets to a truly multidimensional sampling-window setting. 
The proofs of these extensions are guided by the discussion of the $1$-dimensional situation in Shields' monograph~\cite{Shields96}.

\section{SETTINGS}

We consider the $d$-dimensional lattice $\mathbb{Z}^{d}$ and the quadrant $\mathbb{Z}_{+}^{d}$. 
Consider a finite alphabet $\mathcal{A}$, $|\mathcal{A}|<\infty $ and the set of arrays with that alphabet: $\Sigma =\mathcal{A}^{\mathbb{Z}^{d}}$, $\Sigma_{+}=\mathcal{A}^{\mathbb{Z}_{+}^{d}}$. 
We define the set of $n$-words as the set of $n\times\cdots\times n$ arrays $\Sigma^{n}:= \mathcal{A}^{\Lambda _{n}}$ for the $n$-box $\Lambda _{n}:=
\left\{ (i_{1},\ldots ,i_{d})\in \mathbb{Z}_{+}^{d}:0\leq i_{j}\leq n-1,j\in \{1,\ldots ,d\}\right\} $. 
An element $x^n\in\Sigma^n$ has elements $x^n(\mathbf{i})\in\mathcal{A}$ for $\mathbf{i}\in\Lambda_n$. 

Let $\mathfrak{A}^{\mathbb{Z}^{d}}$ denote the $\sigma $-algebra of subsets of $\Sigma $ generated by cylinder sets, i.e., sets of the following kind: 
\begin{equation*}
\lbrack y]:=\left\{ x\in \Sigma :x(\mathbf{i})=y(\mathbf{i}),\mathbf{i\in }%
\Lambda \right\} ,\ \ \ y\in \mathcal{A}^{\Lambda }, |\Lambda|<\infty. 
\end{equation*}%
If $C$ is a subset of $\mathcal{A}^{\Lambda }$, we will use the notation $[C] $ for $\cup _{y\in C}[y]$.
\newline

We denote by $\sigma _{\mathbf{r}}$ the natural lattice translation by the vector $\mathbf{r}\in \mathbb{Z}^{d}$ acting on $\Sigma $ by $\sigma _{\mathbf{r}}x(\mathbf{i}):=x(\mathbf{i}+\mathbf{r})$. 
We use the same notation $\sigma _{\mathbf{r}}$ to denote the induced action on the set $\mathbb{P}$ of probability measures $\nu $ over $(\Sigma ,\mathfrak{A}^{\mathbb{Z}^{d}})$: $\sigma _{\mathbf{r}}\nu (E):=\nu (\sigma _{\mathbf{r}}^{-1}E)$. 
The set of all stationary (translation-invariant) elements of $\mathbb{P}$ is denoted by $\mathbb{P}_{\text{stat}}$, i.e., $\nu \in \mathbb{P}_{\text{stat}}$ if $\sigma_{\mathbf{r}}\nu =\nu $ for each $\mathbf{r}\in \mathbb{Z}^{d}$. 
Those $\nu \in \mathbb{P}_{\text{stat}}$ which cannot be decomposed as a proper convex combination $\nu =\lambda _{1}\nu _{1}+\lambda_{2}\nu _{2},$ with $\nu _{1}\neq \nu \neq \nu _{2}$ and $\nu _{1},\nu_{2}\in \mathbb{P}_{\text{stat}}$ are called \emph{ergodic}. 
The corresponding subset of \ $\mathbb{P}_{\text{stat}}$ is denoted by $\mathbb{P}_{\text{erg}}$. 
Throughout this paper $\mu $ will denote an ergodic $\mathcal{A}$-process on $\Sigma $. By $\nu ^{n}$ we denote the restriction of the measure $\nu $ to the block $\Lambda _{n}$, obtained by the projection $\Pi _{n}:x\in \Sigma \rightarrow x^{n}\in \Sigma ^{n}$ with $x^{n}(\mathbf{i})=x(\mathbf{i}),\mathbf{i}\in \Lambda _{n}$. 
We use the same notation $\Pi _{k}$ to denote the projections from $\Sigma ^{n}$ to $\Sigma^{k},n\geq k$, defined in the same obvious way. 
The measurable map $\Pi _{n}$ transforms the given probability measure $\nu $ to the probability measure denoted by $\nu ^{n}$.
\newline

The entropy rate of a stationary probability measure $\nu $ is defined as limit of the scaled $n$-word entropies: 
\begin{align*}
H(\nu ^{n}):=& -\sum_{x\in \Sigma ^{n}}\nu ^{n}(\{x\})\log \nu ^{n}(\{x\}) \\
h(\nu ):=& \lim_{n\rightarrow \infty }\frac{1}{n^{d}}H(\nu ^{n}).
\end{align*}
Here and in the following we write $\log$ for the dyadic logarithm $\log_{2}$.
\newline 

For a \emph{shift} $\mathbf{p}\in \Lambda _{k}$ we consider the following partition of $\mathbb{Z}^{d}$ into \emph{$k$-blocks}: 
\begin{equation*}
\mathbb{Z}^{d}=\bigcup\limits_{\mathbf{r}\in k\cdot \mathbb{Z}^{d}}(\Lambda
_{k}+\mathbf{r}+\mathbf{p)},
\end{equation*}%
and in general we use the following notation:

The \emph{regular }$k$\emph{-block partitions} of a subset $M\subset\mathbb{Z}^{d}$ are the families of sets defined by 
\begin{equation*}
\mathscr{R}_{M,k}:=\left\{ R_{M,k}(\mathbf{p}): \mathbf{p}\in\Lambda_k
\right\}, \quad R_{M,k}(\mathbf{p}):=\left\{ (\Lambda_k+\mathbf{p}+\mathbf{r})\cap M \right\}_{\mathbf{r}\in k\cdot\mathbb{Z}^{d}}.
\end{equation*}
Clearly, for any $\mathbf{p}$ the elements of $R_{M,k}(\mathbf{p})$ are disjoint and their union gives $M$.
\newline

In the case $M=\Lambda _{n}$, given a sample $x^n\in \Sigma^n$, such a partition yields a \emph{parsing} of $x^n$ in elements of $\mathcal{A}^{(\Lambda _{k}+\mathbf{r}+\mathbf{p)}\cap \Lambda _{n}},\mathbf{r}\in k\cdot \mathbb{Z}^{d}$. 
We call those elements the \emph{words} of the parsing of $x^{n}$ induced by the partition $R_{\Lambda _{n},k}(\mathbf{p})$. With exception of those $\mathbf{r}$, for which $\Lambda _{k}+\mathbf{r}+\mathbf{p}$ crosses the boundary of $\Lambda _{n}$, these are cubic $k$-words. 
Forgetting about their $\mathbf{r}$-position, we may identify $\Pi _{\Lambda _{k}}x\sim \Pi _{\Lambda _{k}+\mathbf{r}}\sigma _{-\mathbf{r}}x\in \mathcal{A}^{\Lambda_{k}+\mathbf{r}}\cong \mathcal{A}^{\Lambda _{k}}$.
\newline

For $k,n\in \mathbb{N}$, $k<n$, any element $x\in \Sigma $ gives rise to a probability distribution, defined by the relative frequency of the different $k$-words in a given parsing of $x_{n}$. 
Let us introduce the following expression for these frequency counts:%
\begin{gather}
Z_{x}^{\mathbf{p},k,n}(a) \;:=\; \sum_{\mathbf{r}\in \times
_{i=1}^{d}\{0,\ldots ,\left\lfloor (n-p_{i})/k\right\rfloor -1\}}\mathbf{1}%
_{[a]}(\sigma _{k\cdot \mathbf{r}+\mathbf{p}}x),\ \ \ \ \  \\
n \in \mathbb{N},\, k\leq n,\, a\in \mathcal{A}^{\Lambda _{k}},\, \mathbf{p}=(p_{1},\ldots ,p_{d})\in \Lambda _{k}.\nonumber
\end{gather}

For regular $k$-block parsings, the \emph{non-overlapping empirical }$k$\emph{-block distribution} generated by $x\in \Sigma $ in the box $\Lambda_{n}$ is the probability distribution on $\Sigma ^{k}$ given by: 
\begin{equation}
\tilde{\mu}_{x}^{k,n}(\{a\}):=\frac{1}{\left\lfloor n/k\right\rfloor ^{d}}%
Z_{x}^{\mathbf{0},k,n}(a)\text{\ \ for }a\in \mathcal{A}^{\Lambda _{k}}.
\end{equation}

Similarly, for any $\mathbf{p}=(p_{1},\ldots ,p_{d})\in \Lambda _{k}$ the
shifted regular $k$-block partition gives a non-overlapping empirical $k$-block distribution: 
\begin{equation}
\tilde{\mu}_{x}^{\mathbf{p},k,n}(\{a\})
:=\frac{1}{\prod_{i=1}^{d}\left%
\lfloor (n-p_{i})/k\right\rfloor }Z_{x}^{\mathbf{p},k,n}(a).
\end{equation}

We will also use the \emph{overlapping empirical }$k$\emph{-block distribution}, in which all $k$-words present in $x$ are considered: 
\begin{equation}
\tilde{\mu}_{x,overl}^{k,n}(\{a\}):=\frac{1}{(n-k+1)^{d}}\sum_{\mathbf{r}\in
\Lambda _{n-k+1}}\mathbf{1}_{[a]}(\sigma _{\mathbf{r}}x)\quad \text{for } a\in \mathcal{A}^{\Lambda _{k}}.  \label{defnonoverlempirical}
\end{equation}

\section{RESULTS}

The main contribution of this paper is the following:

\begin{theorem}[Universally typical sets]
\label{Th1} For any given $0<h_{0}\leq \log|\mathcal{A}|$ there is a sequence of subsets $\left\{ \mathscr{T}_{n}(h_{0})\subset \Sigma ^{n}\right\} _{n}$ such that for all $\mu \in \mathbb{P}_{\text{erg}}$ with $h(\mu )<h_{0}$ the following 
holds:\nopagebreak[4]

\begin{enumerate}
\item[a)] \,\,\, $\underset{n\rightarrow \infty }{\lim}\,\mu^{n}\left( \mathscr{T}_{n}(h_{0})\right) =1$ \quad 
and, in fact, $x^n\in\mathscr{T}_{n}(h_0)$ eventually $\mu$-almost surely.  
\nopagebreak

\item[b)] \,\,\, $\underset{n\rightarrow \infty }{\lim}\,\cfrac{\log |\mathscr{T}_{n}(h_{0})|}{n^{d}}=h_{0}$. \nopagebreak
\end{enumerate}
For each $n$, a possible choice of $\mathscr{T}_n(h_0)$ is the set of arrays with empirical $k$-block distributions of per-site entropies not larger than $h_{0}$, where $k=\left\lfloor \sqrt[d]{\tfrac12 \log _{|\mathcal{A}|}n^{d}}\right\rfloor$. 
\newline

\nopagebreak
\noindent Furthermore, for any sequence $\left\{ \mathscr{U}_{n}\subset\Sigma^n\right%
\}_n $ with $\underset{n\to \infty }{\liminf}\,\frac{1}{n^{d}}\log |\mathscr{U}_{n}|<h_{0}$, there exists a $\mu \in \mathbb{P}_{\text{erg}}$
with $h(\mu )<h_{0}$ which satisfies: \nopagebreak

\begin{enumerate}
\item[c)] \,\,\, $\underset{n\rightarrow \infty }{\liminf}\,\mu^{n} \left( \mathscr{U}_{n} \right) =0$. 
\end{enumerate}
In fact, when  $\underset{{n\to\infty}}{\limsup}\frac{1}{n^d}\log|\mathscr{U}_n|< h_0$, then $x^n\not\in\mathscr{U}_{n}$ eventually $\mu$-almost surely.  
\end{theorem}

The proof of Theorem~\ref{Th1} is based on other assertions following now. 
Although the $1$-dimensional special case of the theorem can be inferred from the existence of universal codes for the class of ergodic processes on $\mathbb{Z}$ and the non-existence of too-good codes, to our knowledge it has not been formulated explicitly before. 
The strategy of our proof is guided by the discussion of $1$-dimensional universal codes contained in~\cite[Theorem~II.1.1, Theorem~II.1.2, and Section~II.3.d]{Shields96}. 
\newline

We start lifting the {\em packing lemma}~\cite[Lemma~I.3.3]{Shields96}. 
We show that if a set of words $C\subset \Sigma ^{m}$ is typical among all $m$-blocks present in a sample $x^{k}\in \Sigma ^{k}$, $k\geq m$, i.e., $C$ has large probability in the overlapping empirical $m$-block distribution, then the sample $x^{k}$ can be parsed into non-overlapping blocks in such a way that nearly all words belong to $C$. 
The following lemma asserts that a parsing with many matchings and only few `holes' can be realized by a regular partition; i.e., $C$ receives large probability in the non-overlapping empirical distribution of some shift of $x$. 

\begin{lemma}[Packing lemma]
\label{packing} For any $0<\delta \leq 1$ let $k$ and $m$ be integers 
satisfying $k\geq d\cdot m/\delta $. Let $C\subset \Sigma ^{m}$ and let $x\in \Sigma $ be such that  $\tilde{\mu}_{x,overl}^{m,k}(C)\geq
1-\delta $. 
Then there is a $\mathbf{p}\in \Lambda _{m}$ such that   
a) $\tilde{\mu}_{x}^{\mathbf{p},m,k}(C)\geq 1-2\delta $, and  
b) $|Z_{x}^{\mathbf{p},m,k}(C)|\geq (1-4\delta )(\left\lfloor \frac{k}{m}\right\rfloor +2)^{d}$.
\end{lemma}

The condition on the array $x$ means that $\sum_{\mathbf{r}\in \Lambda _{k-m+1}}\mathbf{1}_{[C]}(\sigma _{\mathbf{r}}x)\geq (1-\delta )(k-m+1)^{d}$. 
The first statement a) means that there exists a regular $m$-block partition $R_{\Lambda _{k},m}(\mathbf{p})\in \mathscr{R}_{\Lambda _{k},m}$ that parses $x^{k}$ in such a way that at
least a $(1-2\delta )$-fraction of the $m$-words are elements of $C$. 
When $\delta =0$ and $k\geq m$ this statement is trivial. 
The second statement b) implies that at least a $(1-4\delta )$-fraction of the total number of words are elements of $C$ (this total number including non-cubical words at the boundary). 
\newline

\begin{proofi}[Proof of Lemma~\protect\ref{packing}]
Denote by $\Xi $ the set of vectors $\{\mathbf{r}\in \Lambda _{k-m+1}$: $\sigma _{\mathbf{r}}x\text{ is in }[C]\}$. 
For any $\mathbf{p}\in \Lambda_{m}$ denote by $\lambda (\mathbf{p})$ the number of those $\mathbf{r}\in \Xi $ satisfying $\mathbf{r}=\mathbf{p}\text{mod}(m)$. 
Clearly, $\lambda (\mathbf{p})=|Z_{x}^{\mathbf{p},m,k}(C)|$ is the number of cubic blocks in the $p$-shifted regular $m$-block partition of $\Lambda _{k}$ which belong to $C$. 
Then we have $\sum_{\mathbf{r}\in \Lambda _{k-m+1}}\mathbf{1}_{[C]}(\sigma _{\mathbf{r}}x)=\sum_{\mathbf{p}\in \Lambda _{m}}\lambda (\mathbf{p})\geq (1-\delta )(k-m+1)^{d}$, by  assumption. 
Hence, there is at least one $\mathbf{p}^{\prime }\in \Lambda _{m}$ for which $\lambda (\mathbf{p}^{\prime })\geq \frac{(1-\delta )(k-m+1)^{d}}{m^{d}}$. 
It is easy to see that $(1-\delta )\frac{(k-m+1)^{d}}{m^{d}}\geq (1-\delta )\frac{k^{d}-dmk^{d-1}}{m^{d}}\geq (1-\delta )^{2}\frac{k^{d}}{m^{d}}\geq (1-2\delta )\frac{k^{d}}{m^{d}}$.  Since the maximal number of $m$-blocks that can occur in $R_{\Lambda _{k},m}(\mathbf{p}^{\prime })$ is $(\frac{k}{m})^{d}$, this completes the proof of $a)$. 
For $b)$ observe that the total number of partition elements of the regular partition (including the non-cubic at the boundary) is upper bounded by $\left( \left\lfloor \frac{k}{m}\right\rfloor +2\right)^{d}\leq \frac{1}{m^{d}} 
\left( k+2m\right) ^{d}\leq \frac{1}{m^{d}}\left(
k^{d}+(k+2m)^{d-1}2 dm\right) \leq \frac{1}{m^{d}}%
\sum_{j=0}^{d}k^{d-j}(2 dm)^{j}%
\leq \frac{k^{d}}{m^{d}}\frac{1-(2\delta )^{d+1}}{1-2\delta }$. 
Here for the second inequality we used the estimate $1-(d-1)y\leq 1/(1+y)^{d-1},y\geq 0$ and for the third one the estimate $\binom{d-1}{j}\leq d^{j}$. 
On the other hand, from the first part we have $\lambda (\mathbf{p}^{\prime })=|Z_{x}^{\mathbf{p},m,k}(C)| \geq (1-2\delta )\frac{k^{d}}{m^{d}}$ and $1-2\delta \geq \frac{1-4\delta }{1-2\delta }\geq (1-4\delta )\frac{1-(2\delta )^{d+1}}{1-2\delta }$. This completes the proof. 
\hfill
\end{proofi}
\newline

\begin{figure}[]
\begin{center}
\includegraphics[scale=1]{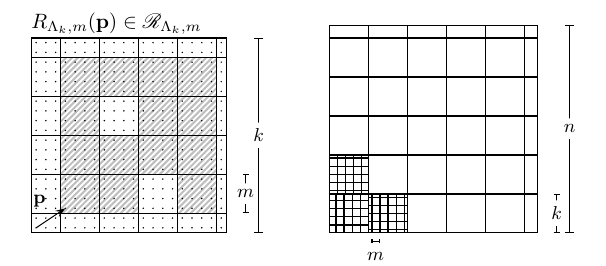}
\end{center}
\par
\par
\begin{center}
\begin{minipage}{11.5cm}
\caption{\small \textbf{Left:} A $\vp$-shifted regular $m$-block parsing of an array $x^k \in \mathcal{T}^{\mu}_k(\delta,m)$, for $d=2$. 
The shaded blocks contain $m$-arrays from  $C^{\mu}_m$ and fill at least a $(1-\delta)$-fraction of the total volume $k^2$. For $k\gg m$ the boundary blocks have a negligible volume. 
\textbf{Right:} A $k$-block parsing of an array $x^n$, giving the empirical distribution $\tilde\mu_x^{k,n}$, and possible regular $m$-block parsings of the resulting $k$-blocks. 
}
\label{regularpartition}
\end{minipage}
\end{center}
\end{figure}

We need two definitions before we continue formulating the results: 
\begin{definition}[Entropy-typical sets]
\label{entropytypicalsets} Let $\delta <\frac{1}{2}$. For some $\mu $ with
entropy rate $h(\mu )$ the \emph{entropy-typical sets} are defined as: 
\begin{equation}
C^{\mu}_{m}(\delta ):=\left\{ x\in \Sigma ^{m}:2^{-m^{d}(h(\mu )+\delta )}\leq \mu
^{m}(\{x\})\leq 2^{-m^{d}(h(\mu )-\delta )}\right\} .
\end{equation}
\end{definition}

We use these sets to define the following \emph{typical sampling sets}. 
See Figure~\ref{regularpartition}. 

\begin{definition}[Typical sampling sets]
\label{typicalsetsalgorithm} For some $\mu $,  $\delta <\tfrac{1}{2}$, and $k\geq m$, we define a \emph{typical sampling set} $\mathcal{T}^{\mu}_{k}(\delta ,m)$ as the set of elements in $\Sigma ^{k}$ that have a regular $m$-block partition such that the resulting words belonging to the $\mu$%
-entropy typical set $C^{\mu}_{m}=C^{\mu}_{m}(\delta )$ contribute at least a $(1-\delta)$-fraction to the (slightly modified) number of partition elements in that regular $m$-block partition. 
\begin{equation*}
\mathcal{T}^{\mu}_{k}(\delta ,m):=\Big\{x\in \Sigma ^{k}:\!\!\!\!\!\!\!\!\!\!\sum 
_{\substack{ \mathbf{r}\in m\cdot \mathbb{Z}^{d}:  \\ (\Lambda _{m}+\mathbf{r%
}+\mathbf{p})\subseteq \Lambda _{k}}}\!\!\!\!\!\!\!\!\mathbf{1}%
_{[C^{\mu}_{m}]}(\sigma _{\mathbf{r}+\mathbf{p}}x)\geq (1-\delta )\left( \frac{k}{m%
}\right) ^{d}\text{ for some }\mathbf{p}\in \Lambda _{m}\Big\}.
\end{equation*}
\end{definition}

We fix some $\alpha >0$ and assume $\delta < {\alpha }/( \log |\mathcal{A}|+1)$. 
In the following we will choose $m$ depending on $k$ such that $m\xrightarrow{k\to\infty}\infty$ and $\lim_{k\rightarrow \infty }\frac{m}{k}=0$. 
As it turns out, a sequence $\mathcal{T}^{\mu}_{k}(\delta ,m)$ satisfying these conditions, denoted $\mathcal{T}_{k}(\alpha)$, is a sequence of `small' libraries from which the realizations of the ergodic process $\mu$ can be constructed asymptotically almost surely. 
Theorem~\ref{Th2} generalizes a previous result by Ornstein and Weiss~\cite[Section~2, Theorem~2]{Ornstein1990} (see~\cite[Theorem~II.3.1]{Shields96}). 

\begin{theorem}
\label{Th2} Let $\mu \in \mathbb{P}_{\text{erg}}$ and $\alpha \in
~(0,\tfrac{1}{2})$. Then:

\begin{enumerate}
\item[a)] For all $k$ larger than some $k_{0}=k_{0}(\alpha )$ there is a set 
$\mathcal{T}_{k}(\alpha)\subset\Sigma ^{k} $ satisfying 
\begin{equation*}
\frac{\log |\mathcal{T}_{k}(\alpha )|}{k^{d}} \leq h(\mu )+\alpha ,
\end{equation*}
and such that for $\mu $-a.e.~$x$ the following holds: 
\begin{equation*}
 \tilde{\mu}_{x}^{k,n}\left( \mathcal{T}_{k}(\alpha )\right)
>1-\alpha ,
\end{equation*}%
for all $n$ and $k$ with $\frac{k}{n}<\varepsilon$ for some $%
\varepsilon=\varepsilon (\alpha )>0$ and $n$ larger than some $n_{0}(x)$. 

\item[b)] Let $\{\tilde{\mathcal{T}}_{k,n}(x)\}_{k,n>0}$ be a family of
double-sequences of subsets of $\Sigma ^{k}$, depending measurably on $x\in
\Sigma $, with cardinality $|\tilde{\mathcal{T}}_{k,n}(x)|\leq 2^{k^{d}(h(\mu
)-\alpha )}$. Then there exists a $k_{1}(\alpha )\geq k_{0}(\alpha )$ and
for $\mu $-a.e.~$x$ there exists an $n_{0}(x)$ such that 
\begin{equation*}
 \tilde{\mu}_{x}^{k,n}(\tilde{\mathcal{T}}_{k,n}(x))\leq \alpha  ,
\end{equation*}%
whenever $k >k_{1}(\alpha )$, $n>n_{0}(x)$, and $2^{k^{d}(h(\mu )+\alpha )}\leq n^{d}$.
\end{enumerate}
\end{theorem}

\vspace{.3cm}

Using Theorem~\ref{Th2} we will prove the following Theorem~\ref{Entropieabsch}, 
which states that the entropy of the non-overlapping empirical distribution of a sample converges almost surely to the true entropy of the process as the size of the parsing blocks grows larger while not exceeding a logarithmic bound with respect to the size of the sampled region. 
In particular, this result describes a procedure to estimate entropies from samples. 
In fact, the inspiring one-dimensional result~\cite[Theorem~II.3.5]{Shields96} is called {\em entropy-estimation theorem}. 
We will use the alternative name {\em empirical-entropy theorem}, referring to its resemblance to the SMB or entropy theorem. 
This result will be a central ingredient in proving the existence of small universally typical libraries (Theorem~\ref{Th1}). 

\begin{theorem}[Empirical-entropy theorem]
\label{Entropieabsch} Let $\mu \in \mathbb{P}_{\text{erg}}$. Then for any
sequence $\{k_{n}\}$ with $k_{n} 
\xrightarrow{n\to\infty}\infty $ and $k_{n}^{d}(h(\mu )+\alpha )\leq \log
n^{d}$ (for some $\alpha >0$) we have 
\begin{equation*}
\lim_{n\rightarrow \infty }\frac{1}{k_{n}^{d}}H(\tilde{\mu}%
_{x}^{k_{n},n})=h(\mu ) ,\quad \mu \text{-a.s.}
\end{equation*}
\end{theorem}

\vspace{.3cm}

This concludes the section of results. Below we provide the proofs.

\section{PROOFS}

\begin{proofi}[Proof of Theorem~\protect\ref{Th2}~\emph{a)}]
We show that the claim holds choosing $\mathcal{T}_{k}(\alpha )$ as typical
sampling sets $\mathcal{T}^{\mu}_{k}(\delta ,m)$ from Definition~\ref{typicalsetsalgorithm} with $\delta <\frac{\alpha }{\log |\mathcal{A}|+1}$, $m\xrightarrow{k\to\infty}\infty$, and $\lim_{k\rightarrow \infty }\frac{m}{k}=0$.
\newline

\emph{Cardinality. } 
We estimate the cardinality of the sets $\mathcal{T}^{\mu}_{k}(\delta ,m)$. 
For a given $m$, there are $m^{d}$ possible values of $\mathbf{p}$. There
are at most $\left( \frac{k}{m}\right) ^{d}$ cubic boxes in any $m$-block
partition of $\Lambda _{k}$. Therefore, the number of choices for the
contents of all blocks which belong to $C^{\mu}_{m}$ is at most $\left\vert
C^{\mu}_{m}\right\vert ^{\left( \frac{k}{m}\right) ^{d}}$. %
By the definition of $\mathcal{T}^{\mu}_{k}(\delta ,m)$, the number of lattice
sites not belonging to the regular partition is at most $\delta k^{d}$. There are $|\mathcal{A}|^{\delta k^{d}}$ possible values for these sites. Let $K=\left\lfloor 
\frac{k}{m}\right\rfloor +2$. The maximal number of blocks in the partition, including non-cubic
ones, is $K^{d}$. For $\frac{m}{k}$ small enough, not more than a $2\delta
\leq \alpha <\frac{1}{2}$ fraction of all these blocks have contents not in $%
C^{\mu}_{m}$. Taking into account that the binomial coefficients $\binom{K}{l}$ do
not decrease in $l$ while $l\leq \frac{1}{2}K$, we get the following bound:%
\begin{eqnarray*}
\left\vert \mathcal{T}^{\mu}_{k}(\delta ,m)\right\vert &\leq &m^{d}\sum_{0\leq
l\leq 2\delta K^{d}}\binom{K^{d}}{l}|\mathcal{A}|^{\delta
k^{d}}|C^{\mu}_{m}|^{\left( \frac{k}{m}\right) ^{d}} \\
&\leq &m^{d}K^{d}\binom{K^{d}}{\left\lfloor \frac{1}{2}K^{d}\right\rfloor }|%
\mathcal{A}|^{\delta k^{d}}|C^{\mu}_{m}|^{\left( \frac{k}{m}\right)^{d}}.
\end{eqnarray*}

We apply Stirling's formula $N!\simeq \sqrt{2\pi N}(\frac{N}{e})^{N}$,
taking into account that the multiplicative error for positive $N$ is
uniformly bounded from below and above. A coarse bound will suffice. In the
following estimate we make use of the relation $\left\vert C^{\mu}_{m}\right\vert
\leq 2^{m^{d}(h(\mu )+\delta )}$, following immediately from the definition
of $C^{\mu}_{m}$. For some positive constants $c,c^{\prime }$, and $c^{\prime
\prime }$ we have 
\begin{eqnarray*}
\log \left\vert \mathcal{T}^{\mu}_{k}(\delta ,m)\right\vert &\leq &\log
\,\,\,cm^{d}K^{d}\left( \frac{K^{d}}{\left\lfloor \frac{1}{2}%
K^{d}\right\rfloor }\right) ^{K^{d}}\sqrt{\frac{K^{d}}{\left\lfloor \frac{1}{%
2}K^{d}\right\rfloor ^{2}}}|\mathcal{A}|^{\delta k^{d}}|C^{\mu}_{m}|^{\left( \frac{%
k}{m}\right) ^{d}} \\
&\leq &\log \,\,\,c^{\prime }m^{d}3^{K^{d}}K^{d/2}|\mathcal{A}|^{\delta
k^{d}}|C^{\mu}_{m}|^{\left( \frac{k}{m}\right) ^{d}} \\
&\leq &\log \,\,\,c^{\prime \prime }k^{d}3^{(\frac{k}{m}+2)^{d}}2^{(h(\mu
)+\delta +\delta \log \left\vert \mathcal{A}\right\vert )k^{d}} \\
&\leq &{k^{d}\left( h(\mu )+\delta (\log \left\vert \mathcal{A}\right\vert
+1)+\frac{2^{d}}{m^{d}}\log 3+\frac{\log k^{d}+\log c^{\prime \prime }}{k^{d}%
}\right) \ .}
\end{eqnarray*}%

In the last line we used $1/m+2/k\leq 2/m$, which holds when $k/m$ is
large enough. 
When $\delta <\frac{\alpha }{\log |\mathcal{A}|+1}$, and $m$ as well as 
$k$ are large enough (depending on $\alpha $), this yields $\log |\mathcal{T}_{k}(\alpha )|\leq k^{d}(h(\mu )+\alpha )$.
\newline

\emph{Probability bound. }
Ornstein and Weiss' extension~\cite{OrWe83} of the SMB theorem shows\footnote{Here, in fact, we only need the convergence in probability~\cite{Kieffer75}, which ensures $\mu
(C^{\mu}_{m})\xrightarrow{m\to\infty}1$.}: 
\begin{equation*}
\lim_{m\rightarrow \infty }- \frac{1}{m^{d}}\log \mu^m (\Pi _{m}x)=h(\mu )\quad
\mu \text{-almost surely} .
\end{equation*}%

Thus, by the definition of $C^{\mu}_{m}$ (Definition~\ref{entropytypicalsets}), there exists an $m_{0}(\delta )$ such that $\mu^{m}\left( C^{\mu}_{m}\right) \geq 1-\delta ^{2}/5$ for all $m\geq m_{0}(\delta )$. We fix such an $m$. The individual ergodic theorem \cite{Lindenstrauss:2001} asserts that the following limit exists for $\mu $-almost every  $x\in \Sigma$: 
$$\lim_{n\rightarrow \infty }\frac{1}{n^{d}}\sum_{r\in \Lambda _{n}}\mathbf{1}_{[C^{\mu}_{m}]}\left( \sigma _{r}x\right) =\int \mathbf{1}_{[C^{\mu}_{m}]}(x)d\mu(x)=\mu^{m}(C^{\mu}_{m}) ,$$  %
and therefore,  
\begin{equation}
\sum_{r\in \Lambda _{n-m+1}}\mathbf{1}_{[C^{\mu}_{m}]}(\sigma _{r}x)
\;\geq\; (1-\delta^{2}/4)(n-m+1)^{d}
\;>\; (1-\delta ^{2}/3)n^{d}  \label{firproperty}
\end{equation}%
holds eventually almost surely, i.e., for $\mu $-almost every $x$, choosing $n$ large enough depending on $x$, $n\geq n_{0}(x)$.

Take an $x\in \Sigma $ and an $n\in \mathbb{Z}_{+}$ for which this is the case and eq.~\eqref{firproperty} is satisfied. Choose a $k$ with $m<k<n$. Consider the unshifted regular $k$-block partition of the $n$-block $\Lambda_{n}$: 
\begin{equation*}
\Lambda _{n}=\bigcup\limits_{\mathbf{r}\in k\cdot \mathbb{Z}^{d}}(\Lambda_{k}+\mathbf{r)}\cap \Lambda _{n}. 
\end{equation*}%
In the following we deduce from eq.~\eqref{firproperty} that if $k/m$ and $n/k$ are large enough,\ at least a $(1-2\delta )$-fraction of the $k$-blocks in this regular $k$-block parsing of $\Pi _{n}x$ (those which count for the empirical distribution $\tilde{\mu}_{x}^{k,n}$) satisfy 
\begin{equation}
\frac{1}{(k-m+1)^{d}}\sum_{\mathbf{s}\in {\Lambda }_{k-m+1}}\mathbf{1}_{[C^{\mu}_{m}]}(\sigma_{\mathbf{s}+\mathbf{r}}x)
\;\geq\; (1-\delta /4) .
\label{property}
\end{equation}%
This is because if more than the specified $2\delta $-fraction of the $k$-blocks had more than a $\delta /4$-fraction of `bad' $m$-blocks, then the total number of `bad' $m$-blocks in $\Pi _{n}x$ would be larger than 
\begin{equation*}
2\delta \left\lfloor \frac{n}{k}\right\rfloor ^{d}\cdot \frac{\delta }{4} (k-m+1)^{d}  \;\geq\;  \frac{\delta ^{2}}{2}\left( \big(1-\frac{k}{n} \big)\big(1-\frac{m}{k}\big)\right)^{d}n^{d} \;>\; \frac{\delta ^{2}}{3}n^{d}, 
\end{equation*}%
for $\frac{k}{n}$ and $\frac{m}{k}$ small enough, contradicting eq.~\eqref{firproperty}. 
While $n$ had to be chosen large enough depending on $x$, we see that $k$ has to be chosen such that $\frac{k}{n}$ and $\frac{m}{k}$ are both small enough. 

By Lemma~\ref{packing}, if ${k\geq 4 dm/\delta }$, the $k$-blocks which satisfy eq.~\eqref{property} have a regular $m$-block partition with at least a $(1-\delta )$-fraction of all partition members in $C^{\mu}_{m}$. 
Hence, at least a $(1-2\delta )$-fraction of all $k$-blocks in $\Lambda _{n}$ counting for the empirical distribution, belong to $\mathcal{T}^{\mu}_{k}(\delta ,m)$. 
For ${\ 2\delta \leq \alpha }$ we get the probability bound: 
\begin{equation}
\tilde{\mu}_{x}^{k,n}\left( \mathcal{T}^{\mu}_{k}(\delta ,m)\right) \geq 1-\alpha .
\label{firstpart}
\end{equation}%
This completes the proof of Theorem~\ref{Th2}~\emph{a)}. \hfill
\end{proofi} 
\newline

\begin{proofi}[Proof of Theorem~\protect\ref{Th2}~\emph{b)}]
The statement is trivial for $h(\mu )=0$. Let $h(\mu )>0$. 
For a fixed $\delta <\alpha $ consider the sets $E_{n}(\delta )$ of all $x$
in $\Sigma $ with  
\begin{equation*}
\tilde{\mu}_{x}^{k,n}(\mathcal{T}_{k}(\delta ))\geq 1-\delta \quad\text{for all } k\geq k_{0}(\delta ),\,2^{k^{d}(h(\mu )+\alpha )}\leq n^{d} ,
\end{equation*}%
where $k_{0}=k_{0}(\delta )$ is chosen large enough as in the first part of the theorem. 
Consider the sets $D_{n}(\alpha ,\delta )$ of all $x$ in $\Sigma $ with 
\begin{equation*}
\tilde{\mu}_{x}^{k,n}(\widetilde{\mathcal{T}}_{k,n}(x))>\alpha \quad\text{for some }k\text{ with }k\geq k_{0}(\delta ),\,2^{k^{d}(h(\mu )+\alpha )}\leq
n^{d}, 
\end{equation*}
and let 
$$F_{n}(\delta ,\alpha )=[C^{\mu}_{n}(\delta )]\cap D_{n}(\alpha
,\delta )\cap E_{n}(\delta ).$$ 

The restriction $a=\Pi _{n}x$ of any $x\in D_{n}(\alpha ,\delta )\cap E_{n}(\delta )$ can be described as follows.

\begin{enumerate}
\item First we specify a $k$ with $k\geq k_{0}(\delta ),2^{k^{d}(h(\mu )+\alpha
)}\leq n^{d}$ as in the definition of $D_{n}(\alpha ,\delta )$.

\item Next, for each of the $\left\lfloor \frac{n}{k}\right\rfloor ^{d}$ blocks
counting for the empirical distribution, we specify whether this block
belongs to $\widetilde{\mathcal{T}}_{k,n}(x)$, to $\mathcal{T}_{k}(\delta
)\setminus \widetilde{\mathcal{T}}_{k,n}(x)$ or to $\Sigma ^{k}\setminus (%
\mathcal{T}_{k}(\delta )\cup \widetilde{\mathcal{T}}_{k,n}(x)) $.

\item Then we specify for each such block its contents, pointing either to a
list containing all elements of $\widetilde{\mathcal{T}}_{k,n}(x)$, or to a
list containing $\mathcal{T}_{k}(\delta )\setminus \widetilde{\mathcal{T}}%
_{k,n}(x)$ or, in the last case, listing all elements of that block.

\item  Finally, we list all boundary elements not covered by the
empirical distribution.
\end{enumerate}

In order to specify $k$ we need at most $\log n$ bits (in fact, much less,
due to the bound on $k$). We need at most $2\left\lfloor \frac{n}{k}%
\right\rfloor ^{d}$ bits to specify which of the cases under {\em 2.}~is valid for
each of the blocks. For {\em 3.}~we need the two lists for the given $k$. This
needs at most $\left( 2^{k^{d}(h(\mu )+\delta )}+2^{k^{d}(h(\mu )-\alpha
)}\right) k^{d}(\log |\mathcal{A}|+1)$ bits. According to the definitions of 
$D_{n}(\alpha ,\delta )$ and $E_{n}(\delta )$, to specify the contents of
all $k$-blocks, we need at most 
\begin{equation*}
\left(\frac{n}{k}+1\right)^{d}k^{d}\left( \alpha (h(\mu )-\alpha )+(1-\alpha )(h(\mu
)+\delta )+\delta (\log |\mathcal{A}|+1)\right) 
\end{equation*}%
bits. For {\em 4.}~we need at most $(n^{d}-\left\lfloor \frac{n}{k}\right\rfloor
^{d}k^{d})(\log |\mathcal{A}|+1)$ bits. Hence the cardinality of $\Pi
_{n}F_{n}(\delta ,\alpha )$ can be estimated by%
\begin{eqnarray*}
&&\log |\Pi _{n}F_{n}(\delta ,\alpha )| \\
&\leq &\log n+2\frac{n^{d}}{k_{1}^{d}(\alpha )} \\
&&+n^{d}\left( n^{-d\left( 1-\frac{h(\mu )+\delta }{h(\mu )+\alpha }\right)
}+n^{-d\left( 1-\frac{h(\mu )-\alpha }{h(\mu )+\alpha }\right) }\right) 
\frac{d\log n}{h(\mu )+\alpha }(\log |\mathcal{A}|+1) \\
&&+n^{d}\left( 1+\frac{1}{n}\sqrt[d]{\frac{d\log n}{(h(\mu )+\alpha )}}%
\right) ^{d}\left( h(\mu )-\alpha ^{2}+\delta (\log |\mathcal{A}|+2)\right) 
\\
&&+n^{d}\left( 1-\left( 1-\frac{1}{n}\sqrt[d]{\frac{d\log n}{h(\mu )+\alpha }%
}\right) ^{d}\right) (\log |\mathcal{A}|+1) \\
&\leq &n^{d}(h(\mu )-\alpha ^{2}/2+\delta (\log |\mathcal{A}|+2))
\end{eqnarray*}%
bits, supposed $n$ is large enough and $k_{1}(\alpha )$ is chosen
sufficiently large. Now, since $\Pi _{n}F_{n}(\delta ,\alpha )\subset
C^{\mu}_{n}(\delta )$, we get%
\begin{equation*}
\mu (F_{n}(\delta ,\alpha ))=\mu ^{n}(\Pi _{n}F_{n}(\delta ,\alpha ))\leq
2^{-n^{d}(\alpha ^{2}/2-\delta (\log |\mathcal{A}|+3))}.
\end{equation*}%
Making $\delta $ small enough from the beginning, the exponent here is
negative. Hence, by the Borel-Cantelli-lemma, only finitely many of the
events $x\in F_{n}(\delta ,\alpha )$ may occur, almost surely. But we know
from the first part of the theorem that $x\in E_{n}(\delta )$ eventually
almost surely (observe that the condition $2^{k^{d}(h(\mu )+\alpha )}\leq n^{d}$
implies $\frac{k}{n}<\varepsilon (\delta )$ as supposed there, for $n$ large
enough). And we know from the Ornstein-Weiss-theorem that $\Pi _{n}x\in
C^{\mu}_{n}(\delta )$ eventually almost surely. Hence $x\in (\Sigma \setminus F_{n}(\delta
,\alpha ))\cap E_{n}(\delta )\cap \lbrack C^{\mu}_{n}(\delta )]\subset \Sigma
\setminus D_{n}(\delta ,\alpha )$ eventually almost surely. 
This is the assertion \emph{b)} of the theorem.
\end{proofi}
\newline

\begin{proofi}[Proof of Theorem~\protect\ref{Entropieabsch}]
The proof follows the ideas of the proof of the $1$-dimensional statement~\cite[Theorem~II.3.5]{Shields96}.
\newline

Let $\alpha <\frac{1}{4}$ and consider the sets ${\mathcal{T}_{k}(\alpha )}$
from Theorem~\ref{Th2}. 
Consider the sets $U_{k,n}(x):=\{a\in \mathcal{T}_{k}(\alpha ):\tilde{\mu}_{x}^{k,n}(a)<2^{-k^{d}(h(\mu )+2\alpha )}\}$. 
Since $|\mathcal{T}_{k}(\alpha )|\leq 2^{k^{d}(h(\mu )+\alpha )}$, also   $\tilde{\mu}_{x}^{k,n}(U_{k,n}(x))\leq 2^{-k^{d}\alpha }$, for any $x$. 

Consider also the sets $V_{k,n}(x):=\{a\in \mathcal{T}_{k}(\alpha ):\tilde{%
\mu}_{x}^{k,n}(a)>2^{-k^{d}(h(\mu )-2\alpha )}\}$. Obviously $|V_{k,n}(x)|\leq 2^{k^{d}(h(\mu )-2\alpha )}$. 
Now, by the second part of Theorem~\ref{Th2}, for $\mu $-almost every $x$ there exists an $n_{0}(x)$ with $\tilde{\mu}_{x}^{k,n}(V_{k,n}(x))\leq 2\alpha$ whenever $n>n_{0}(x)$, $k>k_{1}(2\alpha )$, and $2^{k^{d}(h(\mu )+2\alpha )}\leq n^{d}$. 

We conclude that, for $\mu $-a.e. $x$, the sets $M_{k,n}(x):=\mathcal{T}_{k}(\alpha )\setminus (U_{k,n}(x)\cup V_{k,n}(x))$ satisfy 
\begin{equation*}
\tilde{\mu}_{x}^{k,n}(M_{k,n}(x))
\;\geq\; 1-4\alpha ,
\end{equation*}%
where we assume that $n>n_{0}(x)$, $k>k_{2}(2\alpha)$, $2^{k^{d}(h(\mu )+2\alpha )}\leq
n^{d}$, and $k_{2}(\alpha )\geq k_{1}(\alpha )$ is chosen such that $2^{-k_{2}(\alpha )^{d}\alpha }<\alpha $. 

Consider now the Shannon entropy of the empirical distribution $\tilde\mu^{k,n}_x$, 
\begin{eqnarray}
H(\tilde{\mu}_{x}^{k,n})
&=& -\sum_{a\in \Sigma ^{k}}\tilde{\mu}%
_{x}^{k,n}(a)\log \tilde{\mu}_{x}^{k,n}(a) \nonumber\\
&=& \underbrace{-\sum_{\Sigma ^{k}\setminus M_{k,n}}\!\!\!\!...}_{\Xi
_{k,n}}\,\underbrace{-\sum_{M_{k,n}}...}_{\chi _{k,n}}\;.\label{eq:Shannon}
\end{eqnarray}%
Let $B_{k,n}(x) := \Sigma ^{k}\setminus M_{k,n}(x)$. 
For the first sum in eq.~\eqref{eq:Shannon} an upper bound is given by\footnote{%
Note that $\sum_{a\in B}p(a)\log p(a)\leq p(B)\log |B|-p(B)\log p(B)$.} 
\begin{equation*}
\Xi _{k,n}\leq \tilde{\mu}_{x}^{k,n}(B_{k,n}(x))k^{d}\log |\mathcal{A}|-%
\tilde{\mu}_{x}^{k,n}(B_{k,n}(x))\log \tilde{\mu}_{x}^{k,n}(B_{k,n}(x))%
 , 
\end{equation*}
and hence $\underset{n\rightarrow \infty }{\lim \sup }\frac{1}{k_{n}^{d}}\Xi _{k(n),n}\leq 4\alpha \log |\mathcal{A}|$ holds $\mu $-a.s.~under the theorem's assumptions. 
\newline

For the second sum in eq.~\eqref{eq:Shannon}, note that the elements $a$ from $M_{k,n}(x)$ satisfy 
\begin{equation*}
k^{d}(h(\mu )-2\alpha)\;\leq\; -\log \tilde{\mu}_{x}^{k,n}(a) \;\leq\; k^{d}(h(\mu
)+2\alpha ) ,
\end{equation*}%
and thus 
\begin{gather*}
\frac{1}{k_{n}^{d}}\chi _{k,n}
\;\geq\; \sum_{a\in M_{k,n}(x)}\tilde{\mu}_{x}^{k,n}(a)(h(\mu )-2\alpha )\geq (1-4\alpha )(h(\mu )-2\alpha ) \\
\frac{1}{k_{n}^{d}}\chi _{k,n}
\;\leq\; \sum_{a\in M_{k,n}(x)}\tilde{\mu}_{x}^{k,n}(a)(h(\mu )+2\alpha )\leq h(\mu )+2\alpha .
\end{gather*}
Therefore we have the following holding $\mu $-a.s.:
\begin{eqnarray*}
 (1-4\alpha )(h(\mu )-2\alpha ) 
&\leq &\underset{n\rightarrow \infty }{\lim \inf }\frac{1}{k_{n}^{d}}H( \tilde{\mu}_{x}^{k(n),n}) \\
&\leq &\underset{n\rightarrow \infty }{\lim \sup }\frac{1}{k_{n}^{d}}H( \tilde{\mu}_{x}^{k(n),n}) \\
&\leq &h(\mu )+\alpha (2+4\log |\mathcal{A}|) .
\end{eqnarray*}%

Finally, note that a sequence $k_{n}$ satisfying the two assumptions of the theorem for some $\alpha >0$ in fact satisfies them for any smaller $\alpha $ too. 
This completes the proof. \hfill 
\end{proofi}
\newline

\begin{proofi}[Proof of Theorem~\protect\ref{Th1}] 
When $h_0=\log|\mathcal{A}|$, the first two items are proven by choosing $\mathscr{T}_n(h_0)=\Sigma^n$. In the following we assume  $h_0 < \log|\mathcal{A}|$.  

1. Each $x\in \Sigma $ gives rise to a family of empirical distributions 
$\left\{ \tilde{\mu}_{x}^{k,n}\right\} _{k\leq n}$. 
For each $n$ we define the set $\mathscr{T}_{n}(h_{0})$ as the set of
elements in $\Sigma ^{n}$ having empirical $k$-block entropy per symbol not
larger than $h_{0}$: 
\begin{equation}
\mathscr{T}_{n}(h_{0}):=\Pi _{n}\left\{ x\in \Sigma :H\left( \tilde{\mu}%
_{x}^{k,n}\right) \leq k^{d}h_{0}\right\}. \label{defTscr}
\end{equation}
Here we have to choose $k$ depending on $n$ (how exactly will be specified later). 

The number of all non-overlapping empirical $k$-block distributions in $\Sigma ^{n}$ is upper bounded by $\left( \left( \frac{n}{k}\right)^{d}\right) ^{\left\vert \mathcal{A}\right\vert ^{k^{d}}}$, since $\left\lfloor \frac{n}{k}\right\rfloor ^{d}$ is the maximal count of any particular $k$-block in the parsing of an element of $\Sigma ^{n}$  and $\left\vert \mathcal{A}\right\vert ^{k^{d}}$ is the number of elements in $\Sigma ^{k}$. 

For the number of elements $x^{n}\in \Sigma ^{n}$ with the same empirical distribution ($\tilde{\mu}_{x}^{k,n}$) we find an upper bound which depends only on the entropy of that empirical distribution: 
For a given $n$ with $\lfloor n/k\rfloor =n/k$, we consider the product measure $P=(\tilde{\mu}_{x}^{k,n})^{\otimes (n/k)^{d}}$ on $\Sigma ^{n}$: 
$P(y^{n})=\prod_{\substack{ \mathbf{r}\in k\cdot \mathbb{Z}^{d}  \\ \Lambda
_{k}+\mathbf{r}\subset \Lambda _{n}}}\tilde{\mu}_{x}^{k,n}(\Pi _{k}(\sigma _{\mathbf{r}}y))$, which yields 
\begin{equation}
P(y^{n})=\prod_{a\in \Sigma ^{k}}\left( \tilde{\mu}_{x}^{k,n}(a)\right)
^{(n/k)^{d}\tilde{\mu}_{x}^{k,n}(a)}=2^{-(n/k)^{d}H\left( \tilde{\mu}%
_{x}^{k,n}\right) },\quad \forall y:\tilde{\mu}_{y}^{k,n}=\tilde{\mu}%
_{x}^{k,n},
\end{equation}%
and thus $|\{y\in \Sigma ^{n}:\tilde{\mu}_{y}^{k,n}=\tilde{\mu}%
_{x}^{k,n}\}|\leq 2^{(n/k)H(\tilde{\mu}_{x}^{k,n})}$.

For a general $n:\lfloor n/k\rfloor\neq n/k$, the entries in the positions $%
\Lambda_n\backslash\Lambda_{k\cdot\lfloor n/k\rfloor}$ may be occupied arbitrarily, giving the following bound: 
\begin{equation}
|\{y\in\Sigma^n:\tilde\mu_y^{k,n}=\tilde\mu_x^{k,n}\}|\leq 2^{\lfloor n/k
\rfloor^d H(\tilde\mu_x^{k,n})}\cdot |\mathcal{A}|^{n^d-(n-k)^d}.
\end{equation}

Now we are able to give an upper estimate for the number $|\mathscr{T}%
_{n}(h_{0})|$ of all configurations in $\Lambda _{n}$ which produce an
empirical distribution with entropy at most $k^{d}h_{0}$:%
\begin{eqnarray*}
|\mathscr{T}_{n}(h_{0})| &\leq & 2^{h_{0}k^{d}\left( \frac{n}{k}\right)^{d}}
|\mathcal{A}|^{n^{d}-(n-k)^{d}} \left( \left( \frac{n}{k} \right)
^{d}\right) ^{\left\vert\mathcal{A}\right\vert ^{k^{d}}}, \\
\log |\mathscr{T}_{n}(h_{0})| &\leq &{n^d h_0 +(n^{d}-(n-k)^{d}) \log |%
\mathcal{A}| +\left\vert \mathcal{A}\right\vert ^{k^d} d\log \frac{n}{k} }.
\end{eqnarray*}

Introducing the restriction 
$k^{d}\leq \frac{1}{1+\varepsilon }\log _{|\mathcal{A}|}n^{d}=\frac{\log
n^{d}}{(1+\varepsilon )\log |\mathcal{A}|}$, with $\varepsilon >0$ arbitrary, 
we conclude that $|\mathscr{T}_{n}(h_{0})|\leq 2^{n^{d}h_{0}+o(n^{d})}$
(uniformly in $k$ under the restriction). This yields $\underset{%
n\rightarrow \infty }{\lim \sup }\frac{\log |\mathscr{T}_{n}(h_{0})|}{n^{d}}%
\leq h_{0}$.\newline

2. Next we have to prove that such a sequence of sets, with $k=k(n)$
suitably specified, is asymptotically typical for all $\mu \in \mathbb{P}_{%
\text{erg}}$ with $h(\mu )<h_{0}$. Given any $\mu $ with $h(\mu )<h_{0}$,
Theorem~\ref{Entropieabsch} states that for $\mu $-a.e. $x$ the $k$-block
empirical entropy $\frac{1}{k}H(\tilde{\mu}_{x}^{k,n})$ converges to $h(\mu )$,
provided $k=k(n)$ is a sequence with $k(n)\rightarrow \infty $ and $%
k^d(n)\leq \frac{\log n^{d}}{h(\mu )+\alpha } $, where $\alpha >0$ can be
chosen arbitrarily. %
Since any $\mu$ satisfies $h(\mu )\leq \log |\mathcal{A}|$, choosing $k^{d}(n)\leq \frac{\log
n^{d}}{(1+\varepsilon )\log |\mathcal{A}|}$ with $\varepsilon >0$ yields assertion {\emph{a)}} by the definition of $\mathscr{T}_{n}(h_{0})$, eq.~\eqref{defTscr}.%
\newline

3. Consider a sequence $\{\mathscr{U}_{n}\subset \Sigma ^{n}\}_{n}$ with $%
\liminf_{n\rightarrow \infty }\frac{1}{n^{d}}\log \left\vert \mathscr{U}%
_{n}\right\vert =h_{1}<h_{0}$. One can find an ergodic $\mu $ with $h(\mu
)=h_{2}$ and $h_{1}<h_{2}<h_{0}$. We know that $\mu ^{n}$ is asymptotically
confined to the entropy typical subsets%
\begin{equation*}
C^{\mu}_{n}(\delta )=\left\{ a\in \Sigma ^{n}:2^{-n^{d}(h_{2}+\delta )}\leq \mu
^{n}(\{a\})\leq 2^{-n^{d}(h_{2}-\delta )}\right\} ,
\end{equation*}%
and therefore   
\begin{equation*}
\underset{n\rightarrow \infty }{\lim \inf }\mu (\mathscr{U}_{n})
\;=\;\underset{n\rightarrow \infty }{\lim \inf }\mu (\mathscr{U}_{n}\cap C^{\mu}_{n}(\delta))
\;\leq\; \underset{n\rightarrow \infty }{\lim \inf }|\mathscr{U}_{n}|2^{-n^{d}(h_{2}-\delta)}
\;=\;\underset{n\rightarrow \infty }{\lim }2^{n^{d}(h_{1}-h_{2}+\delta )}.
\end{equation*}%
Choosing $\delta $ small enough this limit is zero. 
The previous analysis, together with the Borel-Cantelli-lemma, shows that on any subsequence with  $\limsup_{n'\to\infty}\frac{1}{n'^d}\log|\mathscr{U}_{n'}|<h_0 $, only finitely many of the events $x^{n'}\in \mathscr{U}_{n'}$ may occur, almost surely. 
This proves \emph{c)}. 
Combining \emph{c)} and \emph{a)}, we get $\underset{{n\rightarrow \infty }}{\liminf}\frac{1}{n^{d}}\log \left\vert \mathscr{T}_{n}(h_{0})\right\vert \geq h_{0}$. In the first part of the proof we showed $\underset{n\rightarrow \infty }{\lim \sup} \frac{1}{n^{d}}%
\log |\mathscr{T}_{n}(h_{0})| \leq h_{0}$. Thus \emph{b)} is verified as well.\hfill
\end{proofi}\newline

\section{CONCLUSIONS}

We prove multidimensional extensions of theoretical results about samplings of ergodic sources which are important in the design of universal source coding schemes. 
Our results provide a truly multidimensional mathematical framework for the optimal compression of multidimensional data. %
We show that the set of $n\times \cdots \times n$ arrays with empirical $k$-block distributions of per-site entropy not larger than $h_{0}$, defined in eq.~\eqref{defTscr}, 
is asymptotically typical for all ergodic $\mathcal{A}$-processes of entropy rate smaller than $h_{0}$, where $k=\left\lfloor \sqrt[d]{c\log _{|\mathcal{A}|}n^{d}}\right\rfloor $, $0<c<1$. In other words, for all $\mathcal{A}$-processes of entropy rate smaller than $h_{0}$, the probability of the corresponding cylinder set tends to $1$ as $n\rightarrow \infty $. 
These sets have a log cardinality of order $n^{d}h_{0}$.

\section*{ACKNOWLEDGEMENT}
\small
We are grateful to an anonymous referee for detailed comments and valuable suggestions. 

\makesubmdate

\makecontacts

\end{document}